
\magnification=1200
\baselineskip=18pt
\tolerance=100000
\overfullrule=0pt

\centerline{\bf UNIFICATION OF THE SOLUBLE TWO-DIMENSIONAL}

\centerline{\bf VECTOR COUPLING MODELS}

\vskip .85truein

\centerline{by}

\vskip .85truein

\centerline{C. R. Hagen}

\centerline{Department of Physics and Astronomy}
\centerline{University of Rochester}
\centerline{Rochester, NY 14627}

\vskip 1truein

\centerline{\bf Abstract}

The general theory of a massless fermion coupled to a massive vector
meson in two dimensions is formulated and solved to obtain the complete
set of Green's functions.  Both vector and axial vector couplings are
included.  In addition  to the boson mass and the two coupling constants,
a coefficient which  denotes a particular current definition is
required for a unique specification of the model.
  The resulting four parameter theory and its solution are shown
to reduce in appropriate limits to all the known soluble models,
including in particular the Schwinger model and its axial vector variant.

\vfill\eject

The fact that the theory of a massless fermion with a current-current
coupling in one space and one time dimension can be solved exactly was
discovered by Thirring [1] some years ago.  The model
 subsequently was solved by
Johnson [2] who realized that an essential ingredient had to be a very
precise definition of the current operator.  He adopted a procedure in
which the current is realized as an average of spacelike and timelike
limits of the product of two field operators.  This averaging process was
 motivated by the need to obtain a covariant result, but introduced the
somewhat undesirable feature of a timelike limit which does not fit
comfortably into a Cauchy initial value formulation.

The most general solution of this model was obtained by the author [3]
using an extension of Schwinger's gauge invariant definition of the
current $j^\mu$.  Specifically, one writes in the case of a charged
fermion $\psi (x)$  coupled to an external field $A_\mu$ [4]
$$j^\mu (x) = \lim_{x \rightarrow x^\prime} {1 \over 2} \ \psi
(x) q \alpha^\mu \exp \left[ iq \int^x_{x^\prime} dx^\prime_\mu
\left( \xi A^\mu - \eta \gamma_5 \epsilon^{\mu \nu} A_\nu \right) \right]
\psi (x^\prime) \eqno(1)$$
where the Dirac matrices $\alpha^0$ and $\alpha^1$ are conveniently taken
to be the unit matrix and
 the Pauli matrix $\sigma_3$, respectively, and the limit is
 taken from a spacelike direction.  The parameters $\xi$ and
 $\eta$ are required
by Lorentz invariance to satisfy the constraint
$$\xi + \eta =1 \quad .$$
For reasons of symmetry it
is convenient
throughout this paper
 to keep both of these parameters in the general formulation since
$\xi =1$ $(\eta =1)$ corresponds to vector (axial vector) conservation
 [5].  In particular the
  Johnson solution can be seen to coincide with the choice $\xi = \eta
 = {1 \over 2}$.

 Perhaps the most well known of the two-dimensional models is the
 Schwinger model [6] which is simply QED for a massless fermion.  Although
 its greatest success was the confounding of the conventional view that
 gauge invariance implied zero mass, it continues to be studied in
  widely varied
 applications.  An extension of the Schwinger model to the massive vector
 meson case was made by Sommerfield [7] using Johnson's current
 definition and by Brown [8] who employed Schwinger's $\xi =1$ limit of
 Eq. (1).  The author [9] showed that the results of refs. 7 and 8 are
 obtained as the $\xi = {1 \over 2}$ and $\xi =1$ limits
  respectively of a formulation
 in which Eq. (1) is used as the current definition.

 A model in which only a single component of the fermion field was
 coupled via the current operator to a massive vector meson was
 subsequently proposed and solved by the author [10].  This model
 (hereafter SCM) has the same Green's functions as one which was
 subsequently proposed by Jackiw and
 Rajaraman [11].  The latter formulation has come to be known as the
 chiral Schwinger model.  However, as has been pointed out [12], the name
 is somewhat unfortunate since the equations of motion are inconsistent
 with zero bare mass for the photon even though the Green's functions are a
 consistent set by virtue of the equivalence to the SCM.

 More recently an attempt has been made [13] to generalize the SCM by
 allowing  an arbitrary admixture of vector and axial vector coupling.  This
 could in principle provide an interpolation between the Schwinger model
 and the SCM.  However, it fails to accommodate a variation  of the
 Schwinger model in which the coupling is through the axial vector
 current as well as the known $\xi$ dependence of the Thirring model and
 the vector meson model of refs. 7-9.  In this paper the task of finding
 the most general  formulation of the soluble two dimensional theories
 is carried out and its solution obtained.  It reduces in all the appropriate
 limits to the known soluble models.

 One begins with the Lagrangian
 $$\eqalign{{\cal L} = {i\over 2}\  \psi \alpha^\mu
  \partial_\mu \psi &+ {1 \over
 4}\  G^{\mu \nu} G_{\mu \nu} - {1 \over 2}\ G^{\mu \nu} (\partial_\mu B_\nu -
 \partial_\nu B_\mu )\cr
 \noalign{\vskip 4pt}%
 & - {1 \over 2} \ \mu^2_0 B^\mu B_\mu + ej^\mu B_\mu + j^\mu A_\mu +
 J^\mu B_\mu\cr}$$
 where $A_\mu$ and $B_\mu$ are external sources
  and $er$ is the axial vector coupling constant.  The
  most general
 current allowed by Lorentz invariance for $e=0$ is
$$\eqalign{j^\mu (x) = \lim_{x \rightarrow x^\prime} &{1 \over 2} \ \psi
(x)  \alpha^\mu  q
 (1+ r \gamma_5) \exp \bigg[ iq \int^x_{x^\prime} dx^\prime_\mu
 \big( \xi A^\mu - \eta \gamma_5 \epsilon^{\mu \nu} A_\nu\cr
 &+r \eta \gamma_5 A^\mu - r \xi \epsilon^{\mu \nu} A_\nu
  \big) \bigg]
\psi (x^\prime) \cr}\eqno(2)$$
 where $\gamma_5 = \alpha^1$.
  By careful application of the action principle and functional
 differentiation techniques a  complete solution of the interacting
 theory can be obtained.

 It is easy to see that the first quantity which must be computed is the
 current correlation function $D^{\mu \nu}$. From general considerations
 one infers that its Fourier transform can be written as
 $$\eqalign{D^{\mu \nu} (p) = \big( &g^{\mu \alpha} + r \epsilon^{\mu
 \alpha}\big) \big( g^{\nu \beta} + r  \epsilon^{\nu \beta}\big) \big[
 D_1 \epsilon_{\alpha \sigma} p^\sigma \epsilon_{\beta \tau} p^\tau\cr
 &+ D_2 p_\alpha p_\beta + D_3 \big( p_\alpha \epsilon_{\beta \sigma}
 p^\sigma + p_\beta \epsilon_{\alpha \sigma} p^\sigma \big) \big]
 \quad . \cr}\eqno(3)$$
 \noindent From ref. 3 it is found that for $e=0,\ D_1 = \xi /\pi p^2,\ D_2 =
 \eta /\pi p^2$, and $D_3 =0$.  The vacuum-to-vacuum transition
 amplitude $<0 \sigma_1 | 0 \sigma_2>$ ($\sigma_1$ and $\sigma_2$ specify
 distinct spacelike surfaces) can be put in the form
 $$\eqalign{<0\sigma_1 |0\sigma_2> \ = &<0\sigma_1 |0\sigma_2>_{_{A=J=0}}
 \exp
 \bigg[ {i \over 2}\ \int J_\mu (x) G^{\mu \nu} (x-x^\prime)J_\nu
 (x^\prime ) dxdx^\prime\bigg]\cr
  \noalign{\vskip 4pt}%
 &\exp
 \bigg[ {i \over 2}\ \int A_\mu (x) D^{\mu \nu} (x-x^\prime)A_\nu
 (x^\prime ) dxdx^\prime\bigg]\cr
 \noalign{\vskip 4pt}%
 &\exp \bigg[ i \int J_\mu (x) M^{\mu \nu} (x-x^\prime) A_\nu (x^\prime
 )dxdx^\prime \bigg] \quad .\cr}$$
 Using the result
 $$<0\sigma_1 |0\sigma_2>\  = \exp \bigg[ -ie \int dx\  {\delta^2 \over
 \delta J^\mu (x) \delta A_\mu (x)} \bigg] <0\sigma_1  | 0\sigma_2
 >_{_{e=o}}$$
 where
 $$\eqalign{<0\sigma_1 |0\sigma_2 >_{_{e=0}}\  =
 \ &\exp \bigg[ {i \over 2} \int
 J_\mu (x) G^{\mu \nu}_0 (x-x^\prime) J_\nu (x^\prime)
 dxdx^\prime \bigg]\cr
  \noalign{\vskip 4pt}%
&\exp \bigg[ {i \over 2} \int A_\mu (x) D^{\mu \nu}_{e=0}
 (x-x^\prime) A_\nu
 (x^\prime) dxdx^\prime \bigg]\cr}$$
 and
 $$G^{\mu \nu}_0 (p) = \left( g^{\mu \nu} + {p^\mu p^\nu \over \mu^2_0}
 \right) {1 \over p^2 + \mu^2_0}\eqno(4)$$
 one obtains the formal result
 $$D = D_{e=0} \left[ 1 - e^2 G_0 D_{e=0} \right]^{-1}
 \quad .\eqno(5)$$
 In this equation (5) all Lorentz indices have been
 suppressed so that
  obtaining
 the actual solution of (5) is much more involved in the general
 case than might otherwise be expected.

 A practical approach to this problem consists of writing
 $$D_i (p) = \sum^\infty_0 D^{(n)}_i \qquad\qquad i=1,2,3$$
 and using Eqs. (3-5) to obtain $D_i^{(n+1)}$ in terms of $D_i^{(n)}$.
 Not surprisingly, in  the
 case of $D_3$ the symmetry in $\mu \leftrightarrow
 \nu$ which characterizes the exact result (3) is not manifest in this
 approach so long as $D_1^{(n)}$ and $D_2^{(n)}$ are unconstrained.
   Thus one simplifies the calculation by imposing that symmetry
 in each order, thereby leading to an expression for $D_3^{(n)}$ in terms
 of $D_1^{(n)}$ and $D_2^{(n)}$.  The consequence of this is that one
 obtains a two dimensional matrix  relation
  (rather than a three dimensional one)
 of the form
 $$D^{(n+1)} = K D^{(n)}$$
 where $D = \pmatrix{D_1\cr
 D_2\cr}$ and
 $$\eqalign{&\bigg[ (r^2 -1) + {p^2 \over \mu^2_0} (\xi r^2 -\eta )\bigg]
 K =\cr
 \noalign{\vskip 8pt}%
 &\left[\matrix{\xi (1-r^2)^2 \bigg( 1+\eta {p^2\over \mu^2_0} \bigg) + 2
 {p^2 r^2 \over \mu^2_0} (r^2 -1) \xi^2 + {p^4r^4 \over \mu^4_0}\xi^2
 &-{r^2 p^4 \xi^2 \over \mu^4_0}\cr
 \noalign{\vskip 8pt}%
{\eta^2 r^2 p^4 \over \mu^4_0} &- \eta (1-r^2)^2 \bigg( 1 + \xi
 {p^2 \over \mu^2_0} \bigg) - 2 \eta^2 (1-r^2)
 {p^2 \over \mu^2_0}
 - {p^4\eta^2 \over \mu^4_0}\cr}\right]\ \ .\cr}$$
 The solution is thus
 $$\pmatrix{D_1\cr
  \noalign{\vskip 6pt}%
 D_2\cr} = (1-K)^{-1} \pmatrix{\xi\cr
  \noalign{\vskip 6pt}%
 \eta\cr}\  {1 \over \pi p^2 }\quad .$$
 Very considerable algebra allows one to solve this formal equation.  As a
 first step it is found that
 $$\det (1-K) = \left[ 1 - {e^2 \over \pi \mu^2_0}\ \left( \xi r^2 + \eta
 \right)\right] \ {p^2 + \mu^2 \over p^2 + \mu^2_0}\eqno(6)$$
 where $\mu$ the physical renormalized mass of the theory is given by
 $$\mu^2 = \mu^2_0 {[1 + {\xi e^2 \over \pi \mu^2_0} (1-r^2)][1-
 {\eta e^2 \over \pi \mu^2_0} (1 -r^2)] \over
 1- {e^2 \over \pi \mu^2_0} (\xi r^2 +\eta)} \quad .$$
 It is encouraging to note that for $r=0$ one obtains
 $$\mu^2 = \mu^2_0 + \xi e^2 / \pi$$
 appropriate to the vector meson model [9]; $r=\mu_0 = \eta =0$
 gives $$\mu^2 = e^2 /\pi$$
 of the Schwinger model; $r= \pm1$ gives
 $$\mu^2 = \mu^2_0 \left[ 1- {e^2 \over \pi \mu^2_0}\right]^{-1}$$
 as in the SCM (after allowing for a trivial difference in the definition
 of  e); and for $e\rightarrow 0$,
  $r \rightarrow \infty$, $er \rightarrow e$, $\xi =0$ gives
 $$\mu^2 = e^2 / \pi$$
 as it must for the axial vector Schwinger model.  The result (6) allows
 one finally to obtain
 $$\eqalign{D_1 (p) &= {\xi \over \pi C_\xi} \bigg[ D(p) +
 {e^2 \xi \over \pi \mu^2_0}\ {C_\eta \over 1-
 {e^2 \over \pi \mu^2_0} (\xi r^2 + \eta)} \Delta (p) \bigg]\cr
  D_2 (p) &= {\eta \over \pi C_\eta} \bigg[ D (p) +
 {e^2 \eta \over \pi \mu^2_0}\ {C_\xi \over 1-
 {e^2 \over \pi \mu^2_0} (\xi r^2 + \eta)} \Delta (p) \bigg]\cr
  D_3 (p) &=  -{e^2 \xi \eta r  \over \pi^2 \mu^2_0}
 {1  \over  1 -
 {e^2 \over \pi \mu^2_0} (\xi r^2 + \eta)} \Delta (p)\cr}$$
 where
 $$\eqalign{C_\xi &= 1 + {\xi e^2 \over \pi \mu^2_0} \ (1 - r^2)\cr
 C_\eta &= 1 - {\eta e^2 \over \pi \mu^2_0} (1-r^2)\cr
 D(p) &= 1 / p^2\cr}$$
 and
 $$\Delta (p) = 1 /(p^2 +  \mu^2) \quad .$$
 One immediate result which follows from $D^{\mu \nu} (p)$ is the equal
 time commutator of the charge density $j^0 (x)$ with the current density
 $j^1(x)$.  It has the form
 $$\eqalign{[j^0 (x) , j^1 (x^\prime )] = &- {i \over \pi}\
 \partial_1 \delta (x-x^\prime) {1 \over C_\xi C_\eta} \bigg\{  1 + r^2\cr
 &+ {e^2 \over \pi \mu^2_0} {(\xi C_\eta + r^2 \eta C_\xi )^2 + r^2 (\xi
 C_\eta + \eta C_\xi)^2\over
 1- {e^2 \over \pi \mu^2_0} (\xi r^2 +\eta)}\bigg\} \quad .\cr}$$

 The vector meson propagator can now be computed in terms of $D^{\mu
 \nu}$ according to
 $$G^{\mu \nu} = G_0^{\mu \nu} + e^2 G^{\mu \alpha}_0 D_{\alpha \beta}
 G^{\beta \nu}_0 \quad .$$
 The result is
 $$\eqalign{G^{\mu \nu} (p) = \bigg( &g^{\mu \nu} + {p^\mu p^\nu \over
 \mu^2 } \bigg) \Delta + {e^2 \over \pi \mu^2_0}\ {1 \over
 C_\xi C_\eta \mu^2_0}\ \bigg\{ \bigg[ r \big( p^\mu \epsilon^{\nu \alpha}
 p_\alpha + p^\nu \epsilon^{\mu \alpha}p_\alpha \big)\cr
 &+ (1+r^2) p^\mu p^\nu \bigg] (D - \Delta) + p^\mu p^\nu \
 {1+ r^2 - {e^2 \over \pi \mu^2_0}\ (\eta + \xi r^4) \over 1 - {e^2 \over
 \pi \mu^2_0}\ (\xi r^2 + \eta)}\  \Delta \bigg\}\quad . \cr}$$
 Finally, the calculation of the bosonic sector is completed by means of
 $$M^{\mu \nu} = eG^{\mu \alpha}_0 D_{\alpha \beta} g^{\beta \nu}$$
 which yields
 $$\eqalign{M^{\mu \nu} (p) = &{e \over \pi \mu^2_0}\ {1 \over \mu^2_0}\
 \bigg\{ {1 \over 1- {e^2 \over \pi \mu^2_0} (\xi r^2 +\eta)} \ p^\mu
 \big[ \big( \xi r^2 + \eta \big) p^\nu + r \epsilon^{\nu \alpha} p_\alpha
 \bigg] \Delta \cr
 &+ \bigg[ \epsilon^{\mu \alpha} p_\alpha \epsilon^{\nu \beta} p_\beta
 \bigg( {\xi \over C_\xi} + {\eta r^2 \over C_\eta} \bigg) + p^\mu p^\nu
 \bigg({\eta \over C_\eta} + {\xi r^2 \over C_\xi}\bigg)\cr
 &+ {r \over C_\xi C_\eta} \ \big( \epsilon^{\mu \alpha} p_\alpha p^\nu +
 \epsilon^{\nu \alpha} p_\alpha p^\mu \big) \bigg] (D - \Delta) \bigg\}
 \quad .\cr}$$

 The fermionic sectors of the model can now be obtained in a fairly
 straightforward fashion.  The $2n$-point functions $G(x_1, \dots x_{2n})$
 are calculated from
 $$<0\sigma_1 |0\sigma_2> G(x_1 . \dots x_{2n}) = \exp
 \bigg[ -ie \int {\delta^2 \over \delta A^\mu (x) \delta J_\mu (x)}\bigg]
 G_{e=0} (x_1 , \dots x_{2n})$$
 using [14]
 $$G_{e=0} (x_1 ,\dots x_{2n}) = \exp \bigg[ i \sum_i q_i \int
 G_0 (x-x_i) (1+ r \gamma_5) \alpha^\mu (x)A_\mu (x) dx \bigg]$$
 where $G_0 (x)$ is defined by
 $$\alpha^\mu {1 \over i} \partial_\mu G_0 (x) = \delta (x) \quad .$$
 It is convenient to cast this into the form
 $$\eqalign{G(x_1,\dots x_{2n} )=&\exp \bigg[ i \sum_i q_i \int
 A^\mu (x) N_\mu (x-x_i)dx\bigg]\cr
 & \exp \bigg[ i \sum_i q_i
 \int J^\mu (x) M_\mu (x-x_i) dx \bigg] G_{0,0,e} (x_1 , \dots
 x_{2n})\quad .\cr}$$
 Tedious calculation yields
 $$\eqalign{-i N_\mu (p) = \bigg\{ &p_\mu \bigg[ {1 \over C_\eta} + r
 \gamma_5 {1 \over C_\xi} \bigg] - \epsilon_{\mu \alpha}p^\alpha \gamma_5
 \bigg[ {1 \over C_\xi} + r \gamma_5 {1 \over C_\eta} \bigg] \bigg\}
 D(p)\cr
 &+ {e^2 \over \pi \mu^2_0} \bigg\{ {1 \over 1- {e^2 \over \pi \mu^2_0}
 (\xi r^2 +\eta)} \bigg[ r^2 p_\mu {1 \over C_\eta} + r p_\mu \gamma_5
 {1 \over C_\xi} - r^2 \epsilon_{\mu \alpha} p^\alpha \gamma_5\cr
 &- \epsilon_{\mu \alpha} p^\alpha r\bigg] - \epsilon_{\mu
 \alpha}p^\alpha \gamma_5 \bigg[ {\xi \over C_\xi} -r \gamma_5 {\eta
 \over C_\eta}\bigg] (1-r^2 )\bigg\} \Delta (p)\cr}$$
 and
 $$\eqalign{M_\mu = &{ie \over \mu^2_0}\ \bigg\{ \bigg[ p_\mu \bigg( {1
 \over C_\eta} + r \gamma_5 {1 \over C_\xi} \bigg) - \epsilon_{\mu
 \alpha} p^\alpha \gamma_5 \bigg( {1 \over C_\xi} +r \gamma_5 {1 \over
 C_\eta} \bigg)\bigg] D(p)\cr
 &+ \bigg[ \epsilon_{\mu \alpha} p^\alpha \gamma_5 \bigg( {1 \over C_\xi}
 + r \gamma_5 {1 \over C_\eta} \bigg) + {e^2 \over \pi \mu^2_0} {1 \over
 1- {e^2 \over \pi \mu^2_0}(\xi r^2 +\eta)} \ rp_\mu \bigg( {r \over
 C_\eta} + \gamma_5 {1 \over C_\xi} \bigg) \bigg] \Delta (p) \quad .\cr}$$
  The solution of the model is completed with the calculation of
  $G_{0,0,e} (x_1 , \dots x_{2n})$.  In this case one finds
  $$\eqalign{G_{0,0,e} (x_1, \dots x_{2n}) = &\exp \bigg\{ {ie^2 \over
  \mu^2_0} \sum_{i,j} q_i q_j \bigg[ (1-r^2) \bigg( {1 \over C_\eta} -
  \gamma_{5i} \gamma_{5j} {1 \over C_\xi}\bigg) D(x_i - x_j)\cr
  &+ {1 \over 1- {e^2 \over \pi \mu^2_0}(\xi r^2 +\eta)} \bigg( r^2
  {C_\xi \over C_\eta} + {C_\eta \over C_\xi} \gamma_{5i} \gamma_{5j} +
  r \big( \gamma_{5i} + \gamma_{5j} \big) \bigg) \Delta (x_i - x_j)\bigg]
  \bigg\}\cr
  &G_0 (x_1 , \dots x_{2n}) \quad ,\cr}$$
  a special case of which gives the two point function
  $$\eqalign{G(x) = &\exp \bigg\{ -i \pi \bigg( {e^2 \over \pi
  \mu^2_0}\bigg)^2 (1-r^2)^2 {1 \over C_\xi C_\eta} [D(x) -D(0) ]\cr
  &-{ie^2 \over \mu^2_0} {1 \over 1- {e^2 \over \pi \mu^2_0} (\xi r^2
  +\eta)} \bigg( r^2 {C_\xi \over C_\eta} + {C_\eta \over C_\xi} +2r
  \gamma_5 \bigg) [\Delta (x) -\Delta (0) ] \bigg\} G_0 (x)
\quad   .\cr}$$

The solution derived here agrees in all particulars with results which
have been obtained for the Thirring model $(e, \mu_0 \rightarrow \infty,\
e/\mu_0$ finite), the vector meson model, the
Schwinger model (vector and axial vector) and the single component model.
It should also be noted that it coincides
 with the model of ref. 13 provided
that $\xi = 1$ (i.e., in the case of the original Schwinger definition of
the current).

Before concluding it is of interest to state the limitations which must
be placed on the results obtained.  Clearly, there exists another
(though somewhat less interesting) case in which the current is coupled to
the derivative of a scalar field.  That set of models can be handled by
identical techniques provided that the free vector meson
 propagator is
replaced by $p_\mu p_\nu / (p^2 + \mu^2_0)$.  Are there any other models
which fall outside the scope of this work?  The answer is certainly yes!
An example of such a model is one in which the left and right chiral
projections of the current operator are each coupled to an independent
external vector potential.  If one subsequently couples the sum of the
two chiral currents to a single vector meson
 with equal coefficients, there is no limit in which
such a system can reduce to the Schwinger model.
 This is despite the fact that
the Lagrangians of the two systems formally appear to be the same.  Thus
the model considered in this paper includes all the \underbar{known}
soluble models but also provides a valuable guide in showing that there
must yet exist a wider class of soluble models which are distinct from these.
Details of the model presented here and the additional inequivalent
extensions which it suggests will be provided in subsequent publications.

This work is supported in part by the U.S. Department of Energy Grant No.
DE-FG-02-91ER40685.

\vfill\eject

\noindent {\bf References}

\medskip

\item{1.} W. Thirring, Ann.  Phys. (N.Y.) {\bf 3}, 91 (1958).

\item{2.} K. Johnson, Nuovo Cimento {\bf 20}, 773 (1961).

\item{3.} C.R. Hagen, Nuovo Cimento {\bf 51B}, 169 (1967).

\item{4.} In order to simplify the final form of the fermionic Green's
functions one uses here a charge matrix
$$q = \pmatrix{0 &-i\cr
i &0\cr}$$
\item{  } which acts in a two dimensional charge space of the
 Hermitian field $\psi (x)$.

\item{5.} It is well to point out here that a significant fraction of the
subsequent literature in this subfield chooses to ascribe the effect of
the $\xi$ parameter to the freedom available in the regularization of a
certain logarithmically divergent integral.  Although the more physical
definition of Eq. (1) is the choice of this paper, identical results can
be obtained using either approach.

\item{6.} J. Schwinger, Phys. Rev. {\bf 128}, 2425 (1962).

\item{7.} C. Sommerfield, Ann.  Phys. {\bf 26}, 1 (1964).

\item{8.} L.S. Brown, Nuovo Cimento {\bf 29}, 617 (1963).

\item{9.} C.R. Hagen, Nuovo Cimento {\bf 51A}, 1033 (1967).

\item{10.} C.R. Hagen, Ann. Phys. (N.Y.) {\bf 81}, 67 (1973).

\item{11.} R. Jackiw and R. Rajaraman, Phys. Rev. Lett. {\bf 54}, 1219
(1985).

\item{12.} C.R. Hagen, Phys. Rev. Lett. {\bf 55}, 2223 (1985).

\item{13.} A. Bassetto, L. Grignolo, and P. Zanca, Phys. Rev. {\bf 50D},
1077 (1994).

\item{14.} In dealing with the fermionic sectors of the model a matrix
with a subscript should be taken to mean that it acts on that particular
index of the $2n$ point function.

\end